\documentclass[a4paper]{jpconf}
\usepackage{graphicx}
\usepackage{cite}
\usepackage{subeqn}
\usepackage[usenames]{color}
\usepackage{ulem} 

\begin{document}
\title{Molecular dynamics of halogenated graphene - hexagonal boron nitride nanoribbons}

\author{George Alexandru Nemnes$^{1,2}$, Camelia Visan$^1$, Dragos~Victor~Anghel$^1$, Andrei Manolescu$^3$}

\address{$^1$Horia Hulubei National Institute for Physics and Nuclear Engineering, 077126, Magurele-Ilfov, Romania}
\address{$^2$University of Bucharest, Faculty of Physics, MDEO Research Center, 077125 Magurele-Ilfov, Romania}
\address{$^3$School of Science and Engineering, Reykjavik University, Menntavegur 1, IS-101 Reykjavik, Iceland}

\ead{alexandru.nemnes@nipne.ro}

\begin{abstract}
The hybrid graphene - hexagonal boron nitride (G-hBN) systems offer new routes in the design of nanoscale electronic devices. Using {\it ab initio} density functional theory calculations we investigate the dynamics of zig-zag nanoribbons a few interatomic distances wide. Several structures are analyzed, namely pristine graphene, hBN and G-hBN systems. By passivating the nanoribbon edges with hydrogen and different halogen atoms, one may tune the electronic and mechanical properties, like the band gap energies and the natural frequencies of vibration. 
\end{abstract}

\section{Introduction}

Motivated by recent developments in achieving highly defined patterns in hybrid graphene - hexagonal boron nitride (G-hBN) materials \cite{song,gong} we investigate the structural, electrical and vibrational properties of halogenated G-hBN nanoribbons. Having a lattice mismatch of less than 2\%, graphene and hBN form an ideal mixture of two dimensional materials with rather different conductive properties: while graphene has a high electrical and thermal conductance, hBN has insulating properties. 

By embedding hBN in graphene one opens the possibility of the field effect control over the active region. During the past few years several devices using G-hBN materials have been proposed, such as field effect transistors \cite{britnell,fiori},
spin filters \cite{nemnes2,nemnes3,visan1}, core-shell nanoflakes \cite{nila}, tunneling double barrier structures \cite{nguyen} and thermoelectric devices \cite{yang,visan2,liu2,vishkayi,nemnes4}.

We analyze here the electronic and vibrational properties of graphene, hBN and mixed G-hBN nanoribbons with the focus on the changes induced by passivation with hydrogen and different halogen atoms of different mass and electronegativity.
Atomic-scale control has already been achieved and well-defined graphene nanoribbons have been produced \cite{lit}.
Our computational methodology combines molecular dynamics (MD) with density functional theory (DFT) at the atomic level. The results indicate the possibility of tuning of the band gap energy and the natural vibration frequency by changing the edge passivating atoms. 

\section{Structures and Methods} \label{sec_SM}

\begin{figure}[t]
\begin{center}
 H \hspace{0.5cm} \includegraphics[height=1.5cm]{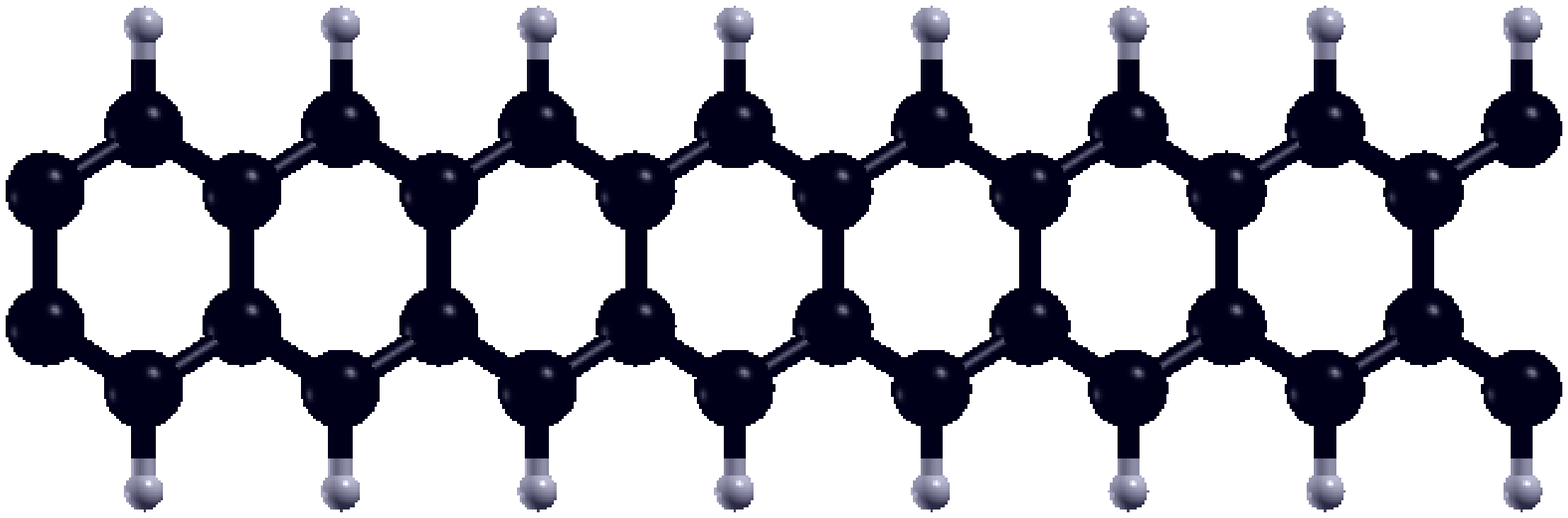}\hspace{0.5cm}
\includegraphics[height=1.5cm]{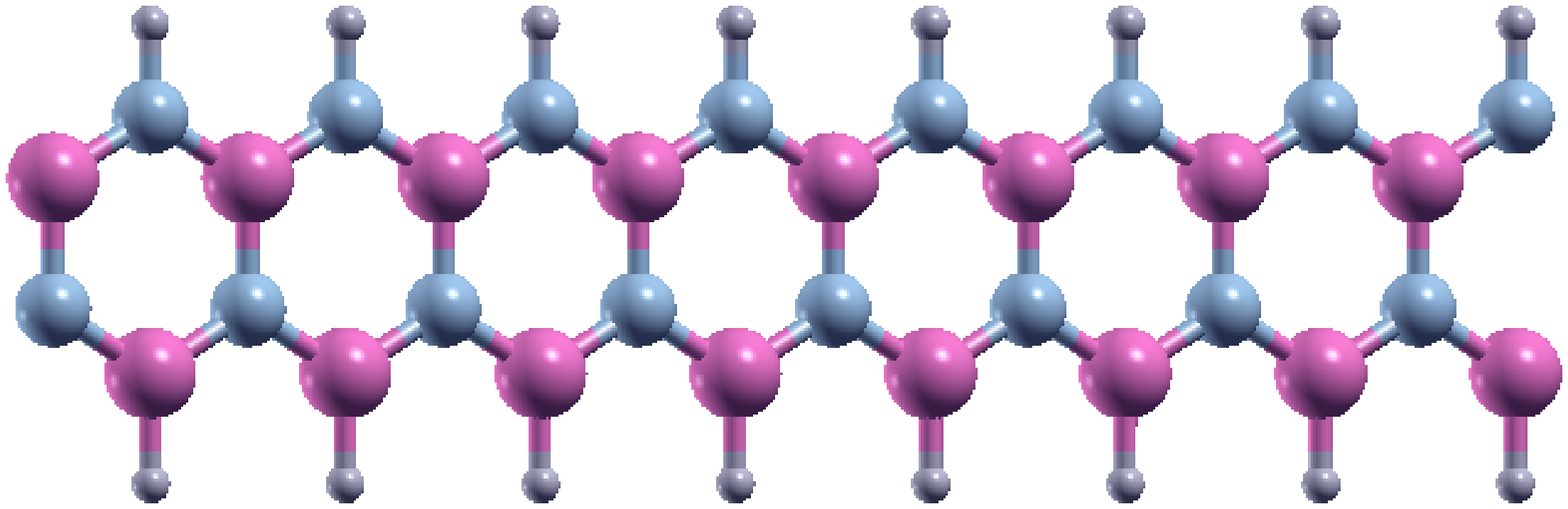}\hspace{0.5cm}
\includegraphics[height=1.5cm]{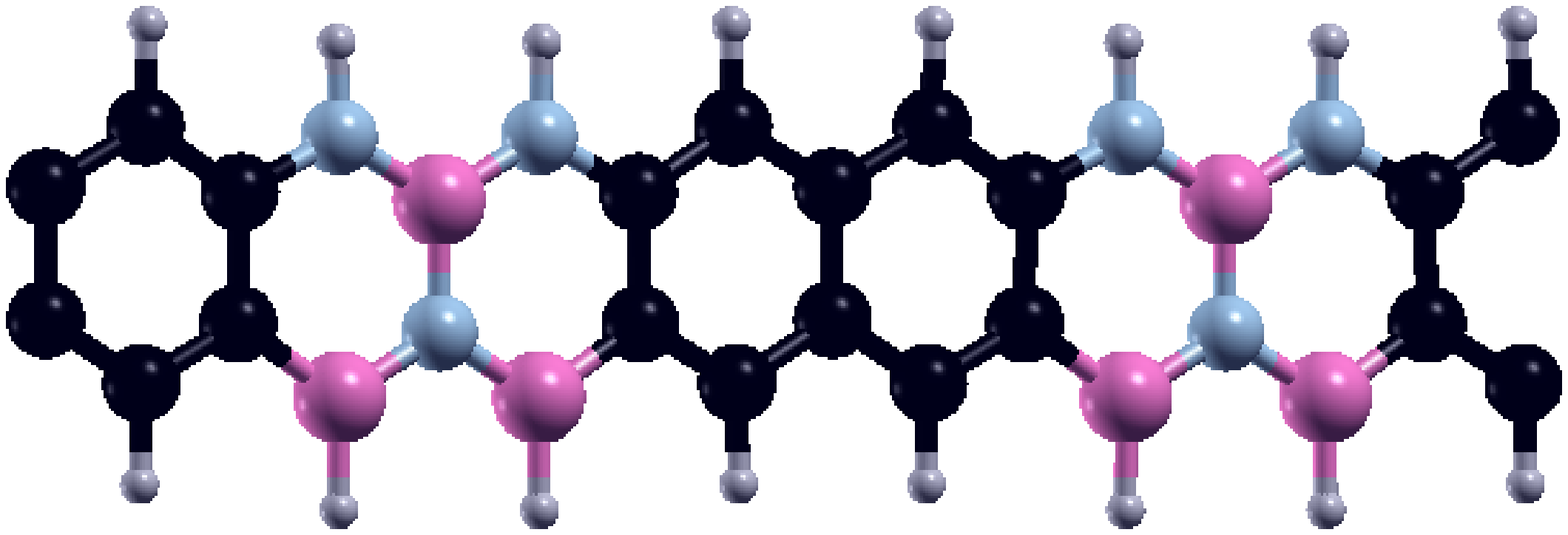}\vspace{0.5cm}\\
 F \hspace{0.5cm} \includegraphics[height=1.7cm]{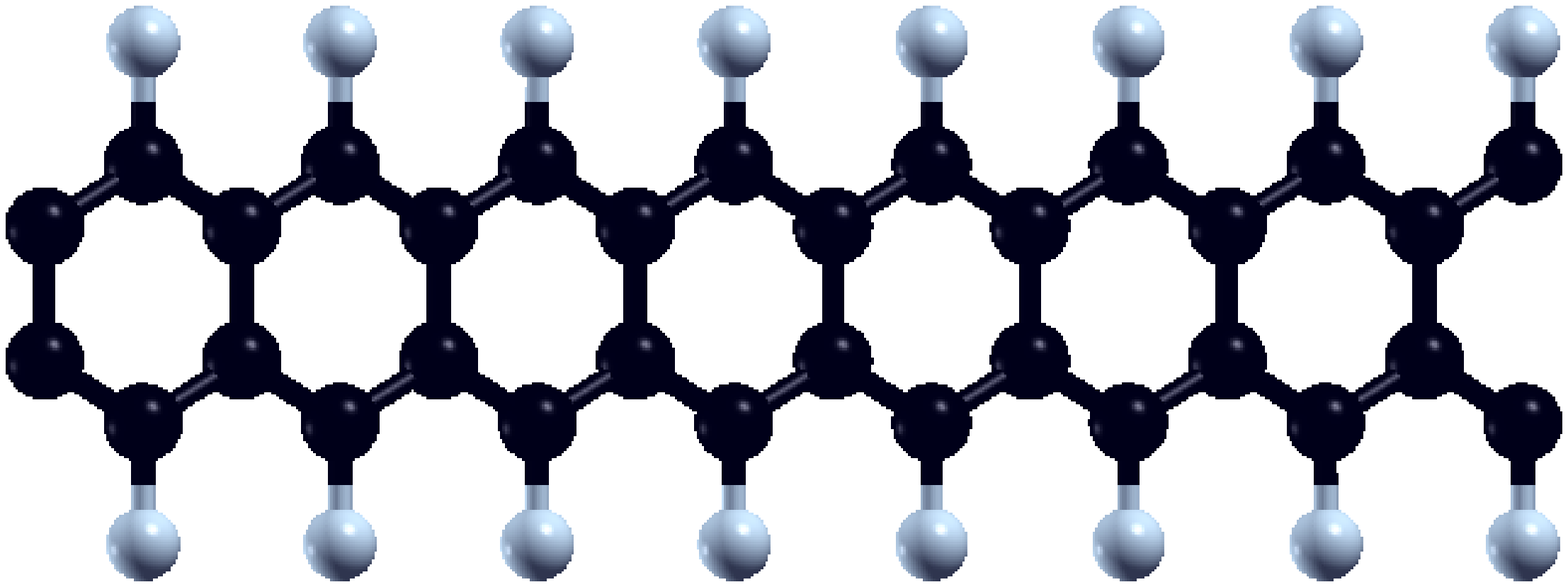}\hspace{0.5cm}
\includegraphics[height=1.7cm]{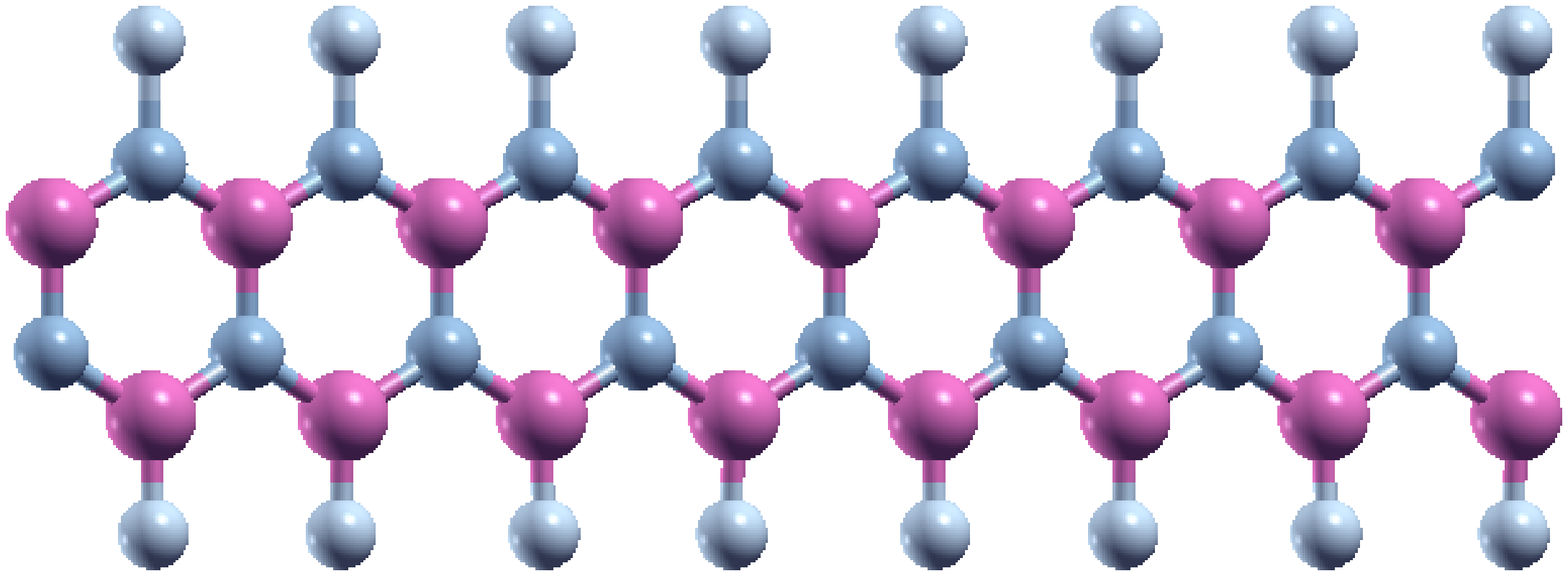}\hspace{0.5cm}
\includegraphics[height=1.7cm]{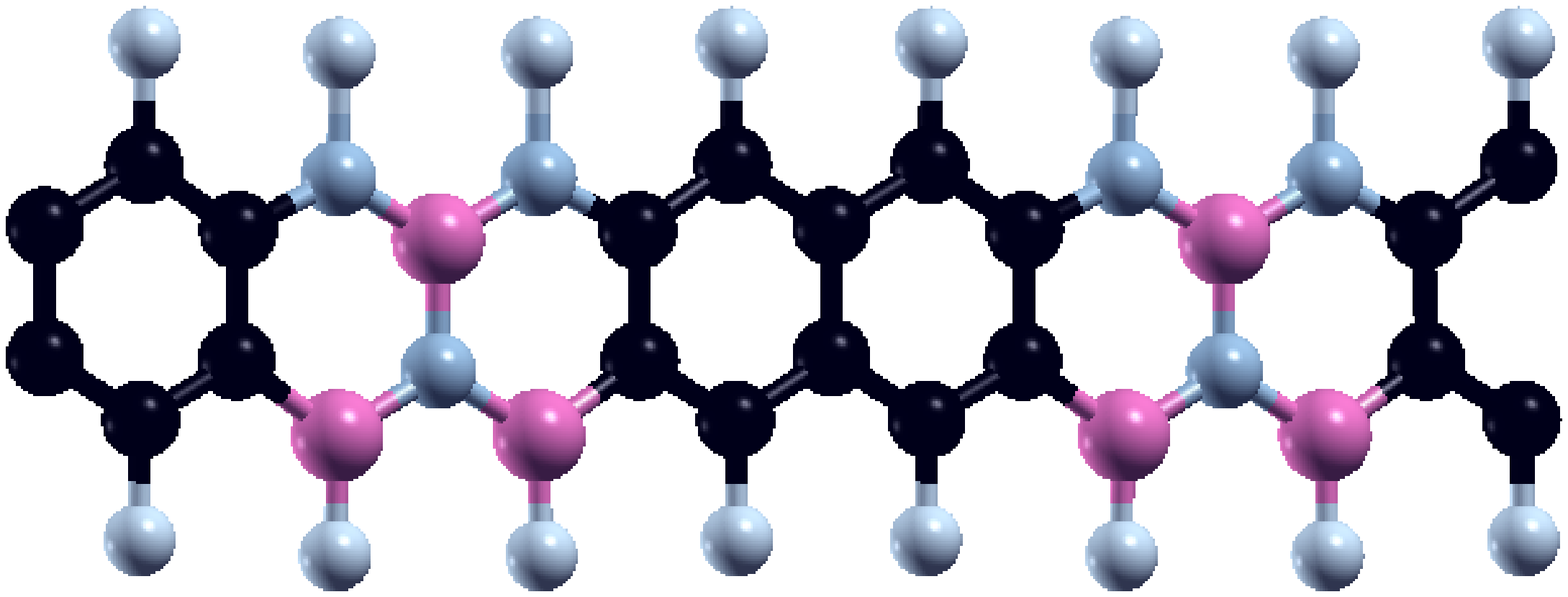}\vspace{0.5cm}\\
Cl \hspace{0.5cm}\includegraphics[height=1.9cm]{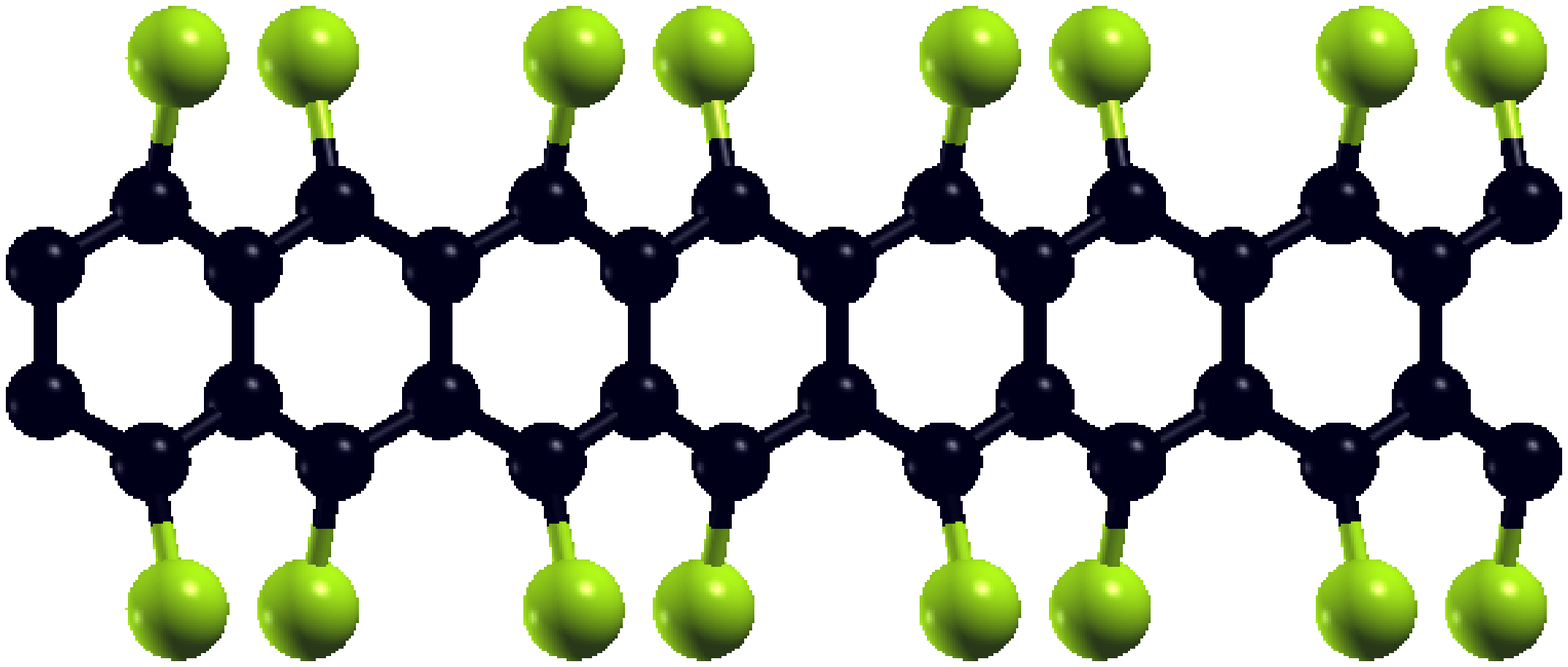}\hspace{0.5cm}
\includegraphics[height=1.9cm]{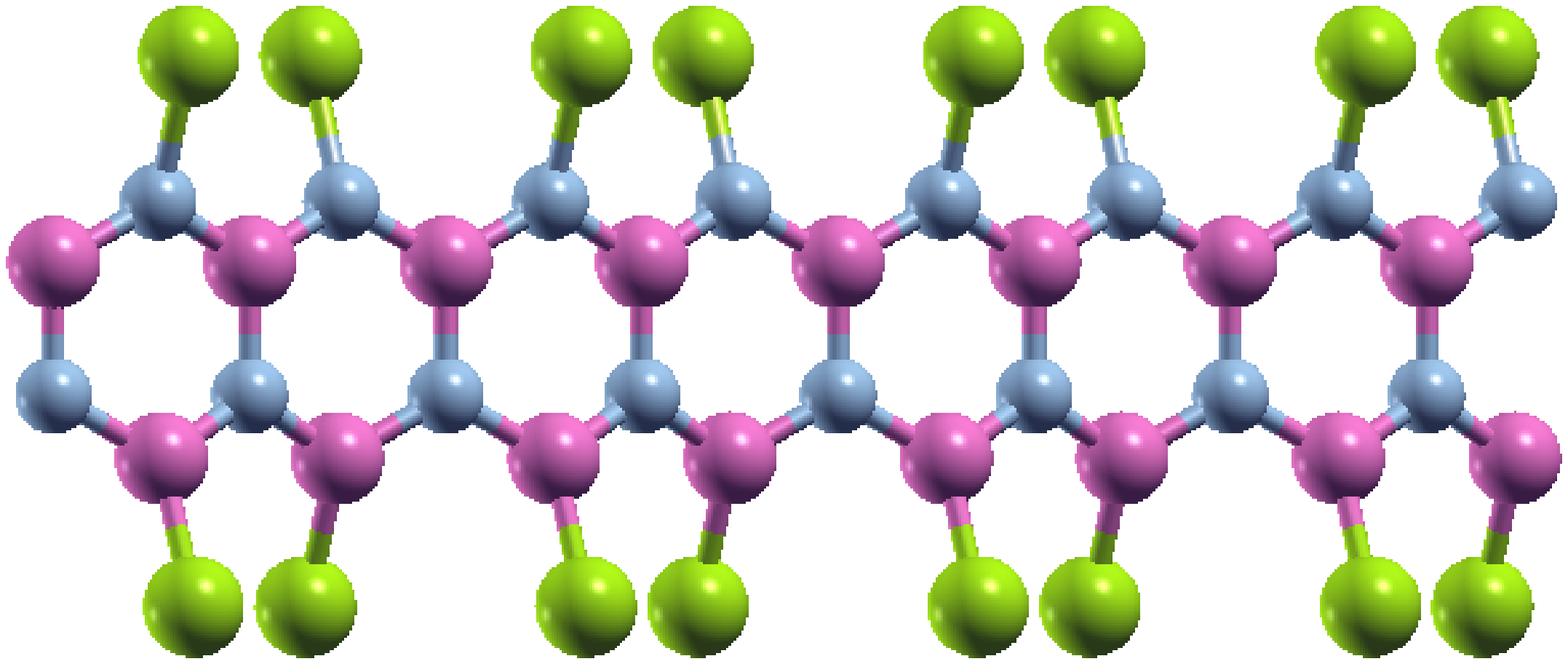}\hspace{0.5cm}
\includegraphics[height=1.9cm]{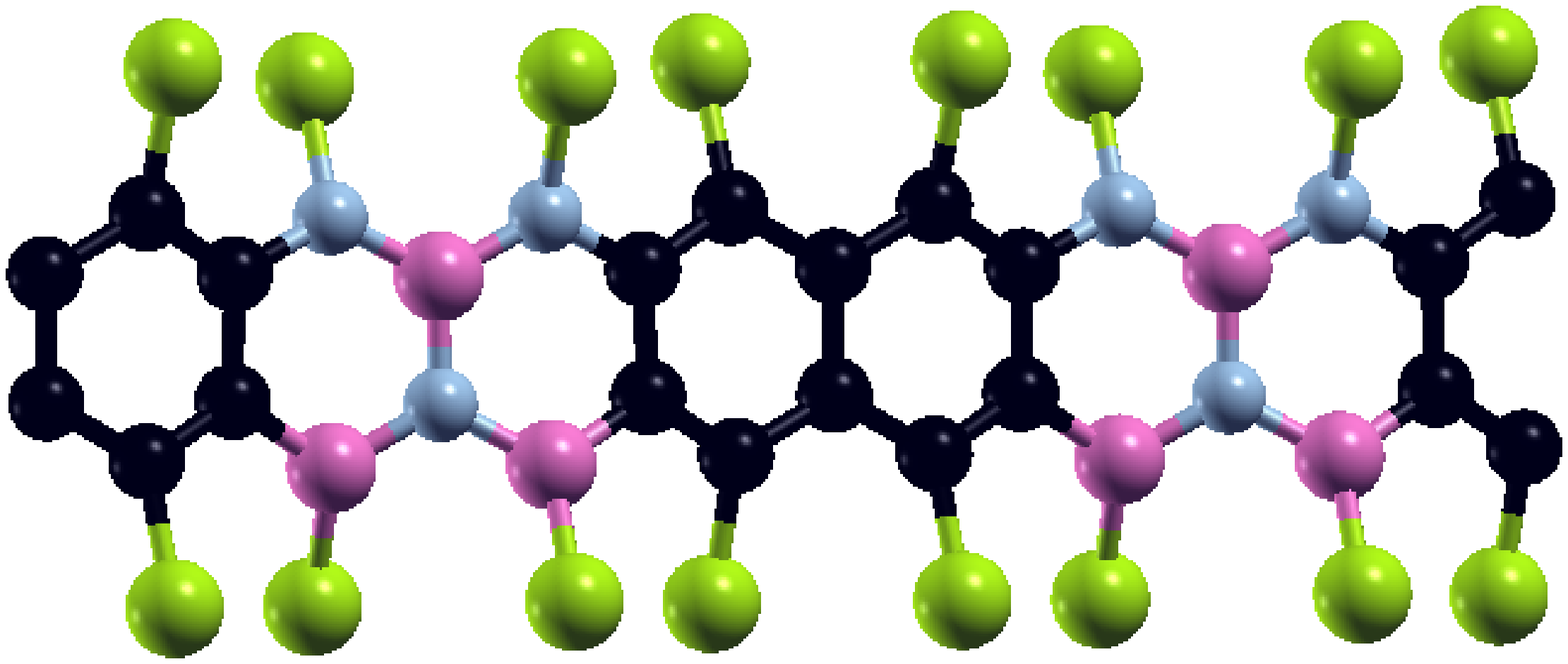}\vspace{0.5cm}\\
Br \hspace{0.5cm}\includegraphics[height=2.05cm]{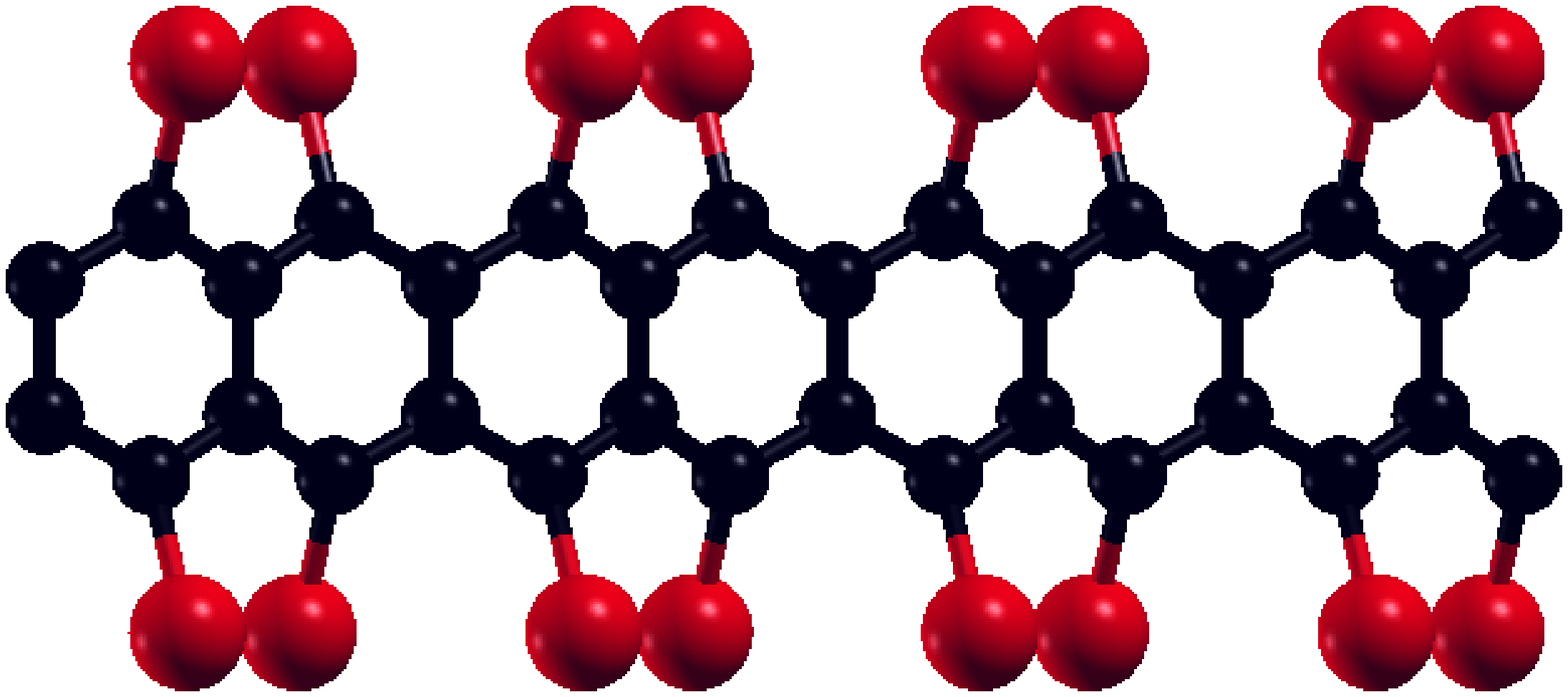}\hspace{0.5cm}
\includegraphics[height=2.05cm]{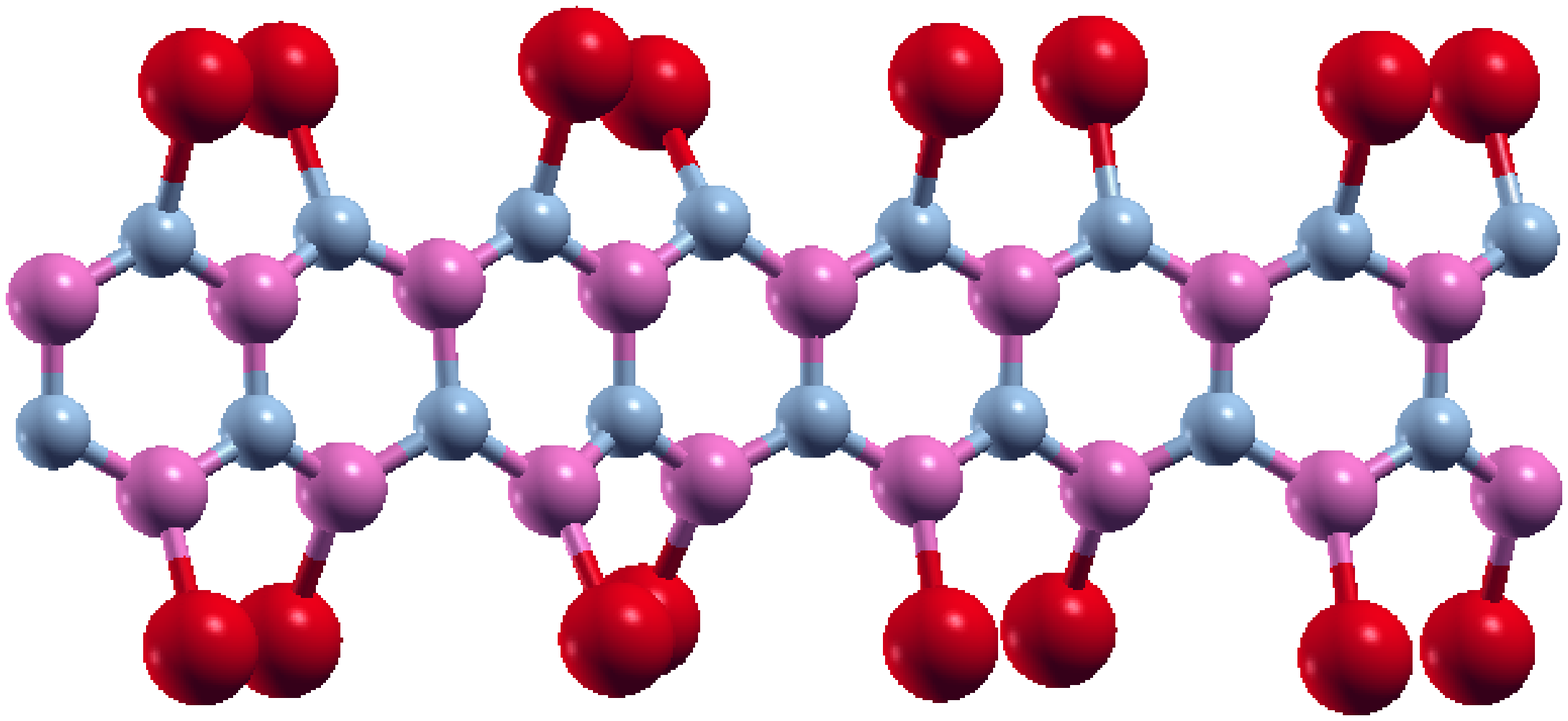}\hspace{0.5cm}
\includegraphics[height=2.05cm]{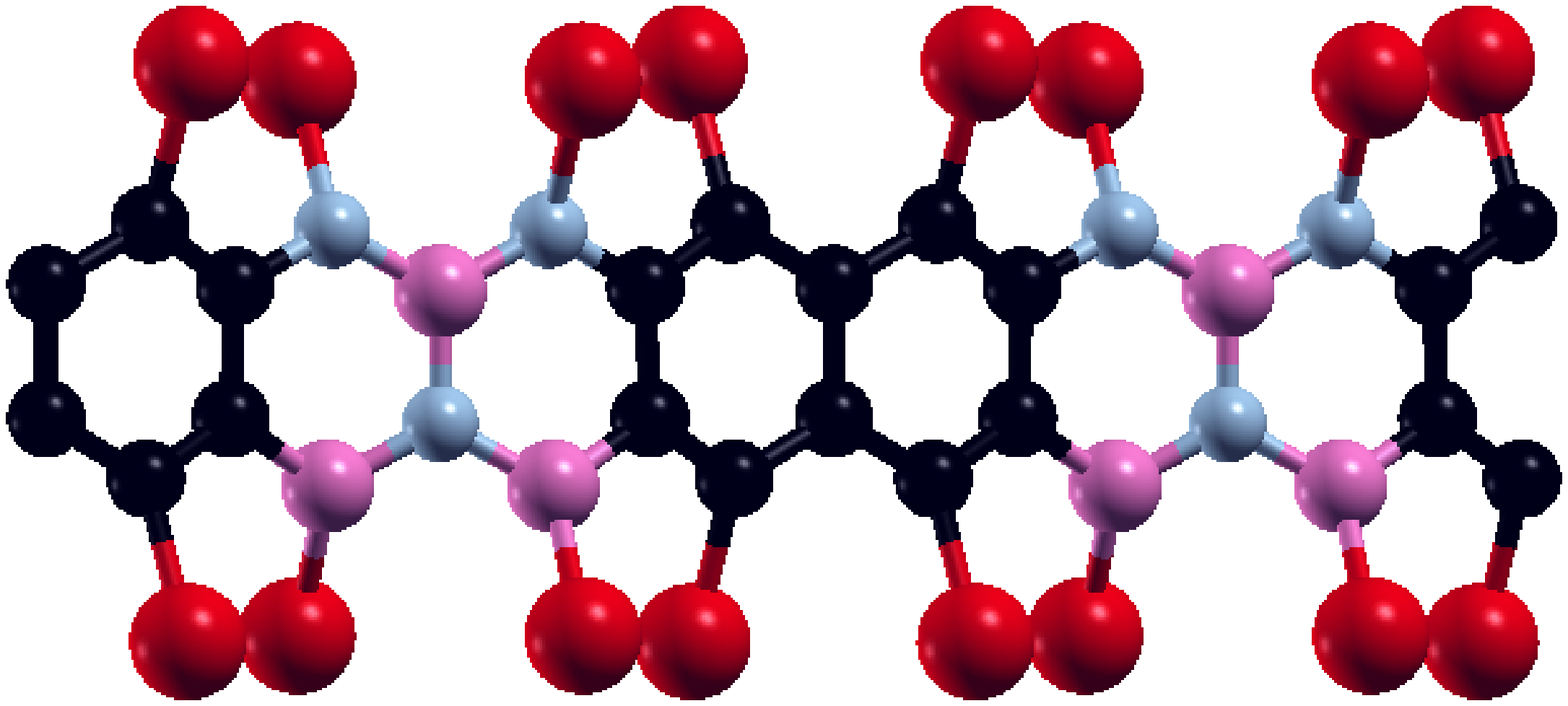}\vspace{0.5cm}\\
 I \hspace{0.5cm}\includegraphics[height=2.2cm]{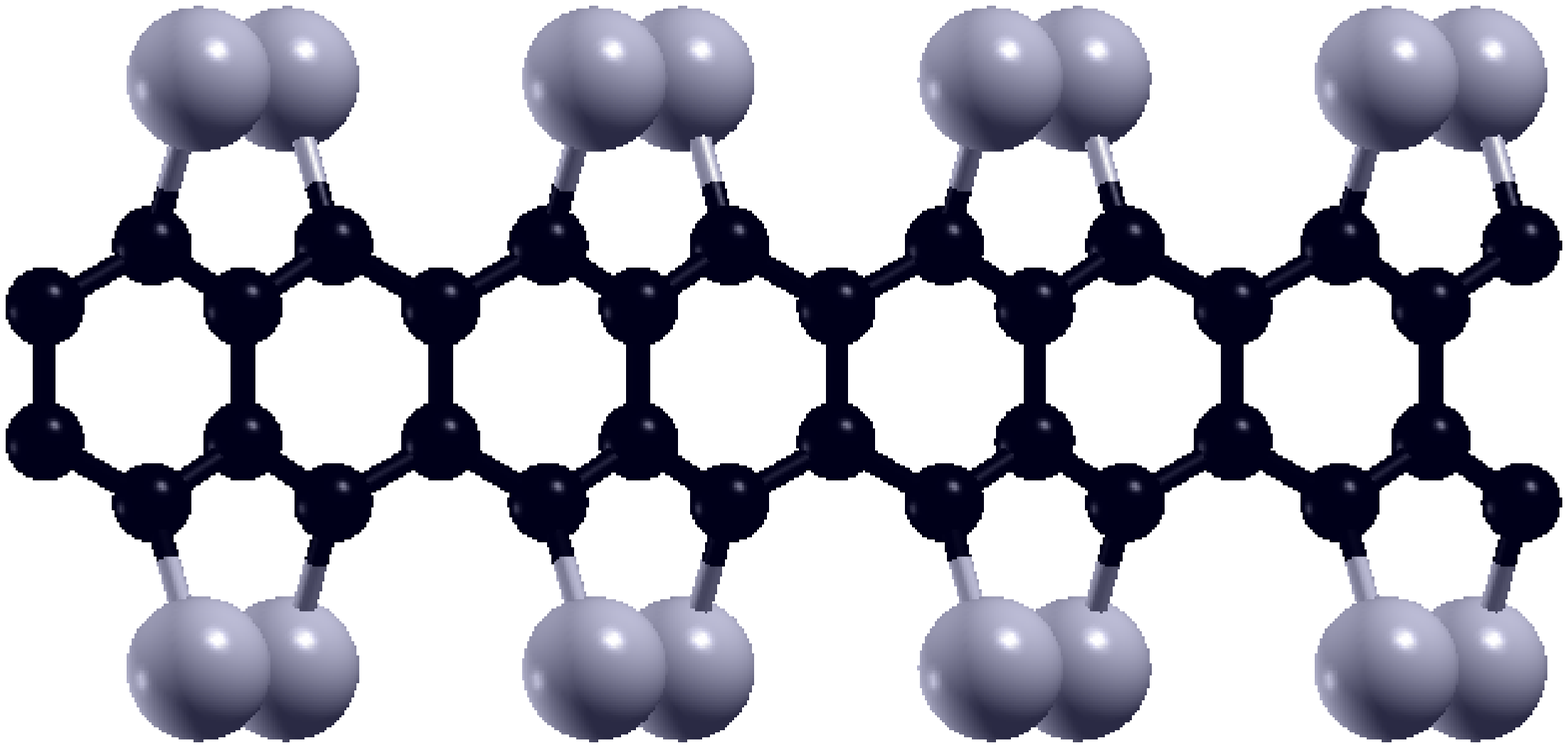}\hspace{0.5cm}
\includegraphics[height=2.2cm]{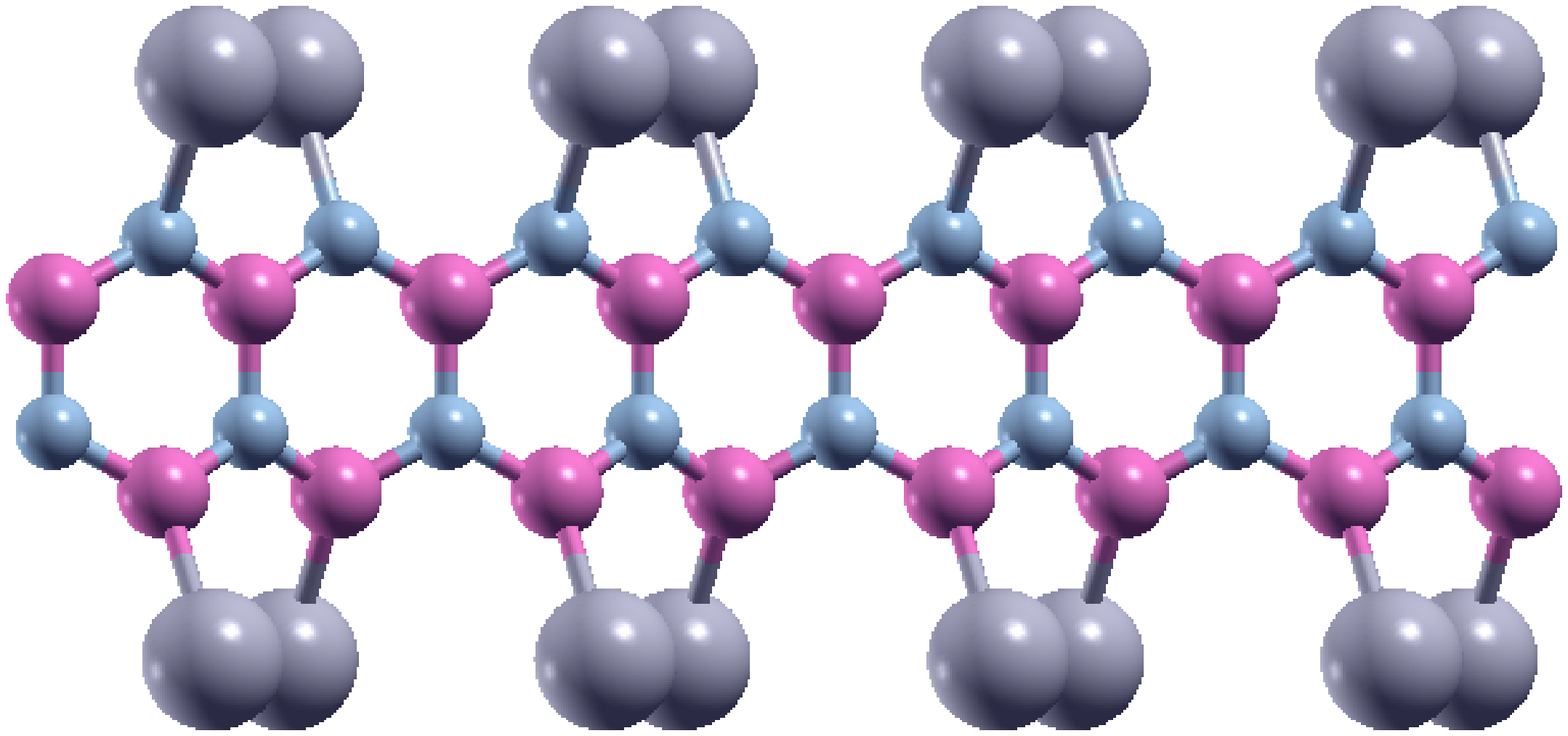}\hspace{0.5cm}
\includegraphics[height=2.2cm]{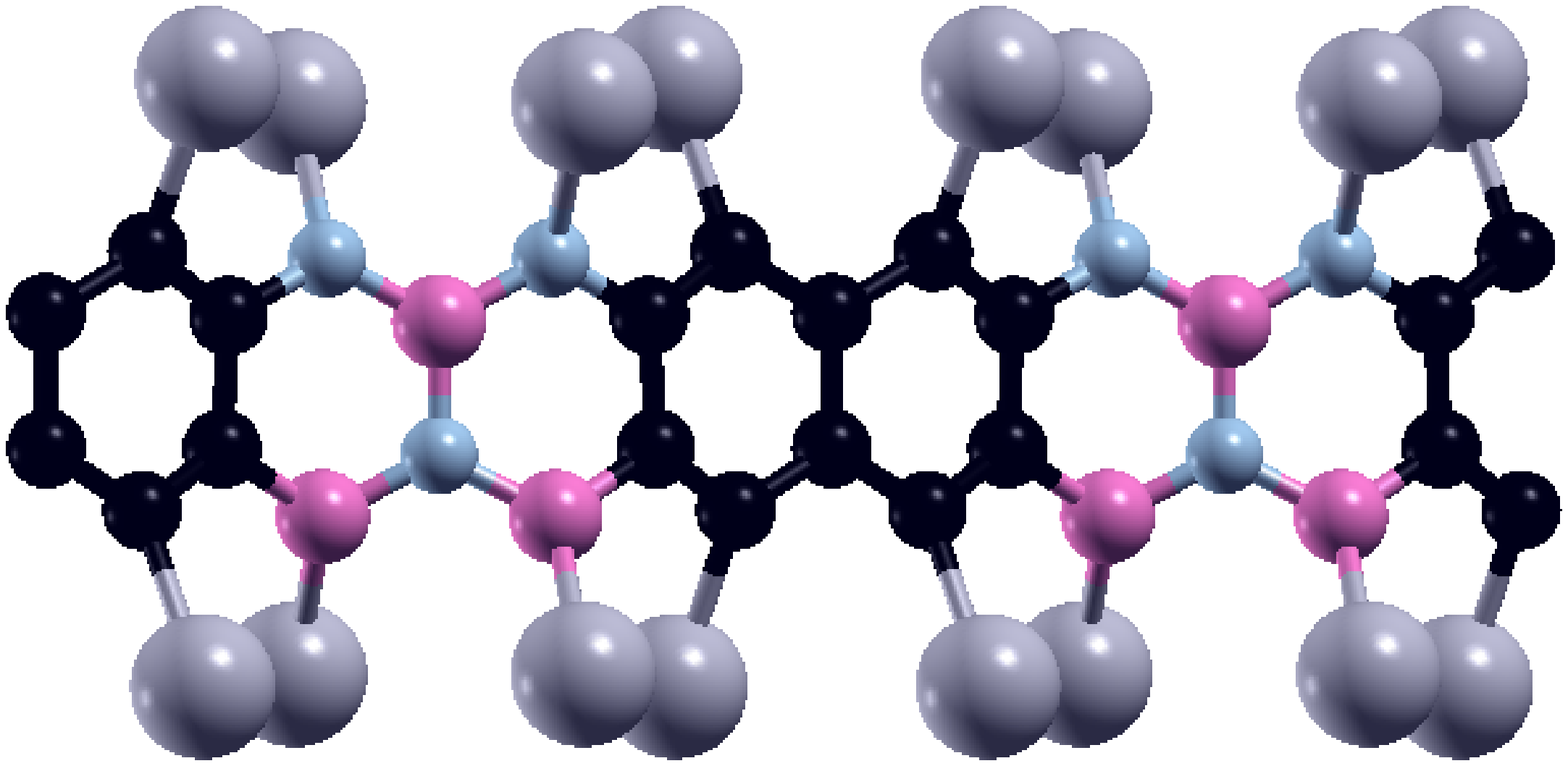}\vspace{0.5cm}\\
\hspace*{0.5cm} G\hspace*{4.5cm} hBN\hspace*{4.0cm} G-hBN\\
\end{center}
\caption{The structures: G, hBN and G-hBN nanoribbons, with the edges passivated by hydrogen (grey) and halogen atoms X = F (light blue), Cl (green), Br (red), I (grey). Boron and nitrogen atoms are depicted in pink and light blue, respectively.}
\label{structures}
\end{figure}

We consider three types of structures: graphene, hBN, and G-hBN nanoribbons with passivated edges, as indicated in Fig.\ \ref{structures}. As the passivating agent we take the hydrogen or one of the halogen atoms X = F, Cl, Br, I. The nanoribbon is oriented along the $x$ direction, with $y$ direction in-plane and $z$ direction perpendicular to the nanoribbon plane.  

The DFT calculations are performed using the SIESTA package \cite{soler}. We employ the local density approximation (LDA) in the parameterization proposed by Ceperley and Alder \cite{ceperley}. The real space grid is fixed by the mesh cutoff energy of 200 Ry and a Monkhorst-Pack grid of $1\times1\times5$ was used. The systems are first relaxed until the residual forces are less than 0.01~eV/\AA. Subsequently, a bent nanoribbon is obtained by linearly displacing the atoms off-equilibrium, in the direction perpendicular to the nanoribbon plane, with the maximum offset $\Delta = 1$\AA\ at the middle of the nanoribbon. Starting from this initial configuration, we perform MD runs of the system coupled to a Nose thermostat, which is kept at a constant temperature of 300~K. The simulation time is 10~ps, with a MD time steps of 1~fs. The simulation time is sufficient to allow the system to perform several oscillations, such that the initial configuration should not affect significantly the results.

\section{Results and discussion}

The structures are first relaxed until the pre-set equilibrium condition is reached. There is a significant structural difference between the systems passivated with H and F, compared to the other three. The H/F passivated nanoribbons retain planar configurations while in the other cases the heavier halogen atoms (Cl, Br, I) are shifted out of the nanoribbon plane.
The minimum energy configurations are obtained for alternating passivating atoms with respect to the nanoribbon plane. There are two possibilities of alternating the atoms on the two edges: in-phase and out-of-phase.
We find that the in-phase structures, in which the passivating atoms (Cl, Br, I) at about the same $x$ coordinate are on the same side of the nanoribbon plane, have lower total energies. For this reason we shall analyze only such structures in the rest of the paper.
The structural data for equilibrium configurations, represented by C-H/X, B-H/X and N-H/X bonding lengths, is listed in Table\ \ref{tabone}. After relaxation, the lengths of the nanoribbon segments are about 20~\AA\ for all structures.

\begin{table}[h]
\caption{\label{tabone}Structural data -- typical bonding lengths [\AA].}

\begin{center}
\lineup
\begin{tabular}{*{6}{l}}
\br                              
  & \hspace*{1cm}H & \hspace*{1cm}F & \hspace*{1cm}Cl & \hspace*{1cm}Br & \hspace*{1cm}I \cr 
\mr
C & \hspace*{1cm}1.11 & \hspace*{1cm}1.32 & \hspace*{1cm}1.71 & \hspace*{1cm}1.88 & \hspace*{1cm}2.07 \cr
B & \hspace*{1cm}1.02 & \hspace*{1cm}1.31 & \hspace*{1cm}1.75 & \hspace*{1cm}1.89 & \hspace*{1cm}2.13 \cr 
N & \hspace*{1cm}1.22 & \hspace*{1cm}1.37 & \hspace*{1cm}1.71 & \hspace*{1cm}1.91 & \hspace*{1cm}2.06 \cr 
\br
\end{tabular}
\end{center}
\end{table}

\begin{figure}[t]
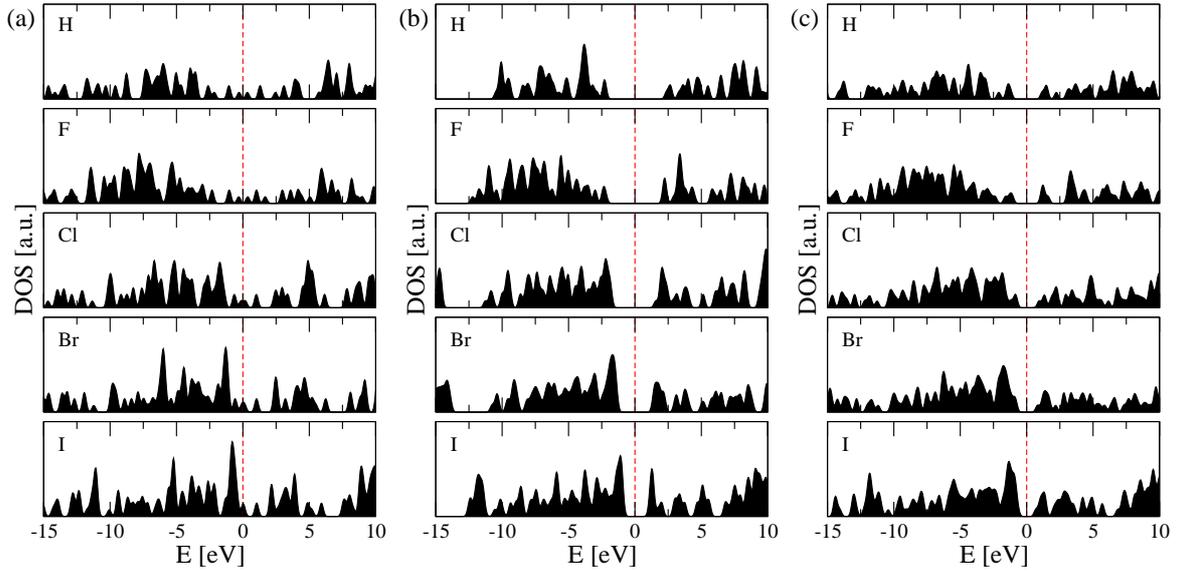

\begin{center}
\includegraphics[height=7.5cm]{G_DOS}\vspace{0.5cm}
\includegraphics[height=7.5cm]{BN_DOS}\vspace{0.5cm}
\includegraphics[height=7.5cm]{G-BN_DOS}
\end{center}
\caption{The density of states of graphene (a), hBN (b) and G-hBN (c) nanoribbons, for different types of passivation. The dashed vertical lines mark the Fermi energy.}
\label{DOS}
\end{figure}

Analyzing the density of states (DOS) of the considered structures, we observe a systematic behavior of the band gap energy. First, the graphene nanoribbons indicate metallic behavior, independent on the passivation type. The DOS just below the Fermi energy typically increases with the halogen atomic number.
Second, the hBN nanoribbons present a wide band gap of about 4~eV for H-passivation. However, the gap decreases monotonically with increasing mass of the halogen atom, reaching a minimum around 1.6~eV, corresponding to iodine passivation. Third, the mixed G-hBN systems have correspondingly smaller gaps than the one observed in hBN systems, with values ranging from around 2~eV (H-passivation) to about 1.1 eV (I-passivation).
Moreover, taking into account that H- and F-passivated systems are planar, while the equilibrium configurations of the other three have out-of-plane halogen atoms, one may also see a resemblance in the DOS spectra within the two groups of systems, (H,F) and (Cl,Br,I), especially around the band gaps.

\begin{figure}[t]
\begin{center}
\includegraphics[height=10cm]{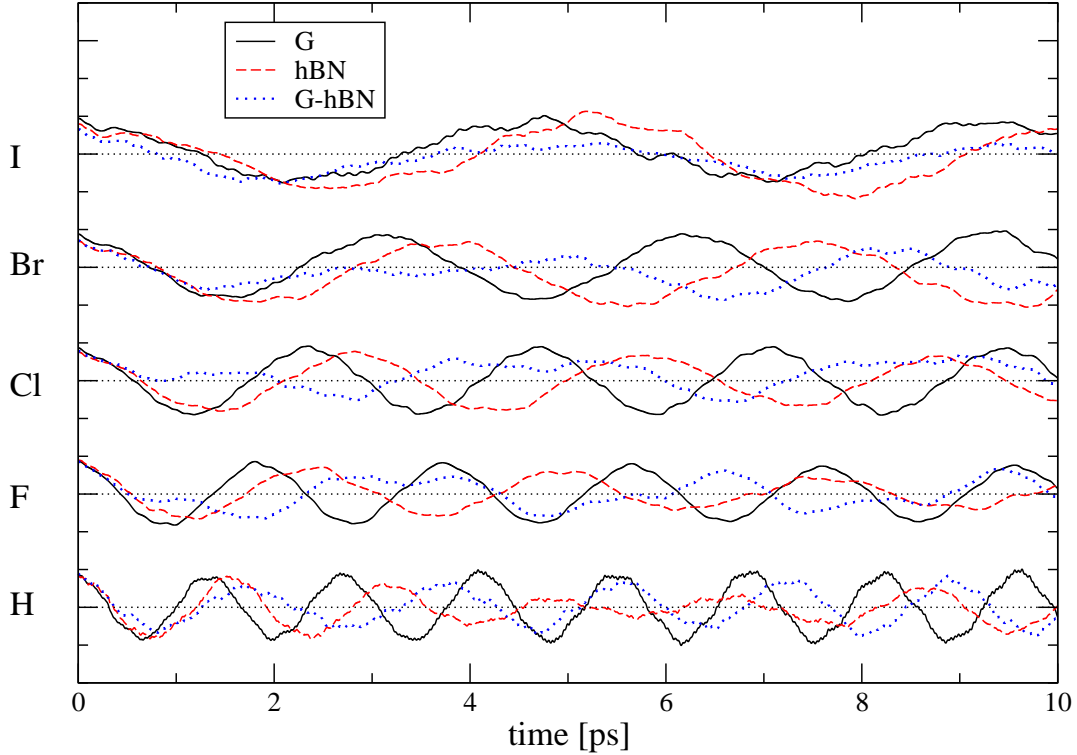}
\end{center}
\caption{Average out-of-plane ($z$) displacement vs. time of the nanoribbon atoms: G (solid lines), hBN (dashed lines) and G-hBN (dotted lines). The spacing between two consecutive dotted lines is 1.5~\AA.}
\label{zm}
\end{figure}

Next, the dynamical properties are investigated.
After the equilibration, the nanoribbons are stretched out of equilibrium by displacing the atoms in the $z$ direction by a distance proportional to their distance to the nearest end.
In this way, the maximum displacement of 1~\AA\ takes place in the middle, whereas the atoms at the extremities of the nanoribbon in the $x$ direction are not moved.
After stretching, the constraints are removed and the nanoribbon starts to oscillate.
The natural frequencies of graphene nanoribbons range from 0.74 THz (H-passivation) down to 0.22 THz (I-passivation). A similar trend is found for hBN and G-hBN structures. Comparing hBN with graphene systems with the same passivation, the oscillating period increases systematically, with frequencies in the range 0.19 - 0.63 THz, which is consistent with other studies where it was found that h-BN sheet is a less stiff material as compared to graphene \cite{singh}. 
The less homogeneous G-hBN system enables more complex oscillating modes. 

\begin{figure}[t]
\begin{center}
\includegraphics[height=10cm]{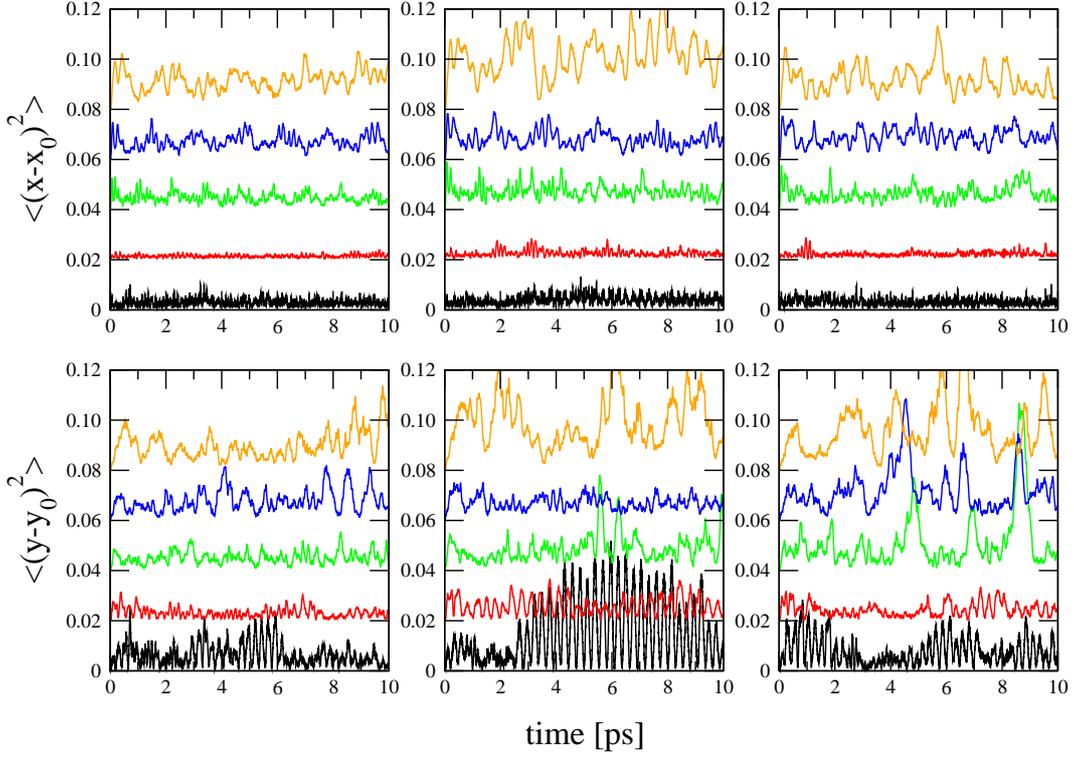}
\end{center}
\caption{Averages of squared displacements [\AA$^2$] along $x$- and $y$ coordinates, with respect to equilibrium positions, for G, hBN, and G-hBN from left to right. The data sets represent H, F, Cl, Br and I passivated systems from bottom to top in each graph, each graph being shifted vertically by 0.02~\AA$^2$ for clarity.}
\label{xy}
\end{figure}

The atomic motion is further analyzed using the averaged squared displacement $\langle(x-x_0)^2\rangle$ and $\langle(y-y_0)^2\rangle$ positions with respect to the equilibrium positions, $x_0$ and $y_0$. The average is taken over all atoms at each time step. Figure\ \ref{xy} shows that all three systems (G, hBN, G-hBN) have roughly the same behavior with respect to changing the passivating atoms: the fluorinated systems indicate the overall smallest average displacements, which can be explained by the planar, more stable configuration. The hydrogenated samples are also planar, however the much smaller mass of H introduces larger fluctuations.
The other three halogenated structures (X=Cl,Br,I) are non-planar and they introduce larger fluctuations, which may potentially impact the transport properties and noise.

\section{Conclusion}

We investigated by {\it ab initio} DFT calculations the electronic and vibrational properties of graphene, hBN and G-hBN nanoribbons, with different types of passivation. By changing the halogen atoms, the physical properties may be tuned: the band gaps of hBN may be decreased by using halogen atoms with higher atomic number; the oscillating period is found to be in the low THz regime and it changes systematically with the mass of the border atoms; the higher stiffness of graphene, compared to hBN is evidenced by higher oscillation frequency.
The type of the passivation proves to be essential for adjusting both electronic and vibrational properties and this may provide a route for optimizing mixed graphene - boron nitride nanostructures. \\

\vspace{1cm}

{\bf Acknowledgments}\\
This work was supported by the National Authority for Scientific Research and Innovation (ANCSI) under grant PN16420202.


\section*{References}
\begin{thebibliography}{99}

\bibitem{song}
Li Song, Lijie Ci, Hao Lu et al., Nanolett. 10, 3209 (2010).

\bibitem{gong}
Yongji Gong, Gang Shi, Zhuhua Zhang et al., Nature Communications 5, 3193 (2014).

\bibitem{britnell}
L. Britnell, R. V. Gorbachev, A. K. Geim, L. A. Ponomarenko, A. Mishchenko, M. T. Greenaway, T. M. Fromhold, K. S. Novoselov, L. Eaves,
Nature Commun. 4, 1794 (2013)

\bibitem{fiori}
G. Fiori, A. Betti, S. Bruzzone and G. Iannaccone,
ACS Nano \textbf{6}, 2642 (2012)

\bibitem{nemnes2} 
G. A. Nemnes, 
J. Nanomater. \textbf{2012}, 748639 (2012).

\bibitem{nemnes3} 
G. A. Nemnes and S. Antohe, 
Mater. Sci. Eng. B \textbf{178}, 1347 (2013).

\bibitem{visan1} 
Camelia Visan, 
Rom. Rep. Phys. \textbf{66}, 983 (2014).

\bibitem{nila}
A. A. Nila, G. A. Nemnes, A. Manolescu,
Rom. J. Phys. 60, 696 (2015).

\bibitem{nguyen}
V. H. Nguyen, F. Mazzamuto, A. Bournel and P. Dollfus, 
J. Phys. D: Appl. Phys. \textbf{45}, 325104 (2012)

\bibitem{yang}
K. Yang, Y. Chen, R. D’Agosta, Y. Xie, J. Zhong, and A.
Rubio, Phys. Rev. B \textbf{86}, 045425 (2012)

\bibitem{visan2} 
Camelia Visan, 
J. Electron. Mater. \textbf{43}, 3470 (2014)

\bibitem{liu2} 
Y. S. Liu, W. Q. Zhou, J. F. Feng, X. F. Wang, 
Chem. Phys. Lett. \textbf{625}, 14 (2015)

\bibitem{vishkayi}
S. I. Vishkayi, M. B. Tagani, H. R. Soleimani,
J. Phys. D: Appl. Phys. \textbf{48}, 235304 (2015) 

\bibitem{nemnes4}
G. A. Nemnes and Camelia Visan 
EPJ Web of Conferences \textbf{108}, 02045 (2016)

\bibitem{lit}
J. van der Lit,	M. P. Boneschanscher, D. Vanmaekelbergh, M. Ij\"as, A. Uppstu, M. Ervasti, A. Harju, P. Liljeroth and I. Swart,
Nat. Commun. 4, 2023 (2013)

\bibitem{soler}
J. M. Soler, E. Artacho, J. D. Gale, A. Garca, J. Junquera, P.
Ordejon, and D. Sanchez-Portal, 
J. Phys. Cond. Mater. \textbf{14}, 2745 (2002)

\bibitem{ceperley}
D.M. Ceperley, B.J. Alder, 
Phys. Rev. Lett. \textbf{45}, 566 (1980)

\bibitem{singh}
S. K. Singh, M. Neek-Amal, S. Costamagna, and F. M. Peeters,
Phys. Rev. B \textbf{87}, 184106 (2013)

\end{thebibliography}
\end{document}